\documentclass[%
 reprint,
 amsmath,amssymb,
 aps,
]{revtex4-2}

\usepackage{graphicx,xcolor}
\usepackage{dcolumn}
\usepackage{bm}
\usepackage{float}
\usepackage{makecell}
\usepackage{amsmath}
\usepackage{csquotes}

\begin{document}

\preprint{APS/123-QED}

\title{On the instabilities of intrinsic thermoacoustic modes in a thermoacoustic waveguide with anechoic terminations}

\author{Haitian Hao}
\email[]{Corresponding author: haoh@purdue.edu}
\affiliation{School of Mechanical Engineering, Purdue University, West Lafayette, Indiana 47907, USA}
\author{Fabio Semperlotti}
\email[]{fsemperl@purdue.edu}
\affiliation{School of Mechanical Engineering, Purdue University, West Lafayette, Indiana 47907, USA}


\begin{abstract}
A recent study \textit{[H. Hao and F. Semperlotti, Phys. Rev. B 104, 104303 (2021)]} investigated the dynamic behavior of an infinite one-dimensional (1D) thermoacoustic waveguide (TAWG) and illustrated its ability to sustain nonreciprocal and near zero-index sound propagation behavior; these properties can be very beneficial in the design of acoustic devices, including acoustic diodes, amplifiers, and cloaks. Nevertheless, it is critical to realize that when this concept is implemented in a finite-length waveguide, dynamic instabilities may occur and either drastically reduce or completely hinder the ability of the TAWG to control and manipulate sound. In this work, we uncover and investigate the occurrence of either evanescent or intrinsic thermoacoustic (ITA) modes in a 1D TAWG with anechoic terminations. The stability analysis clearly distinguishes these two types of evanescent modes and highlights their different origin rooted in either acoustic or thermoviscous effects. Numerical results reveal that ITA modes in anechoic-terminated TAWG are strictly connected to the acoustic-driven evanescent modes, and evolve towards unstable modes as the TA coupling strength is increased. This study may have important implications for the practical design of novel acoustic manipulating devices enabled by TA coupling elements. The conclusions drawn in this study may also shed lights on the effective suppression of instabilities in TAWGs.
\end{abstract}

\maketitle

\section{Introduction}
During the past couple of decades, acoustic metamaterials have seen a significant development motivated by their unique abilities for sound manipulation \cite{MaG, Cummer2016,Zangeneh-Nejad, Liao}. Acoustic metamaterials are artificial structures designed to achieve unique effective properties and wave propagation capabilities such as, for example, negative or zero index properties \cite{HuangHH, FangN, ZhuHF2017}, asymmetric transmission \cite{ShenC, LiY2017, Gu}, anomalous refraction \cite{ZhuHF2016, LiY2014, ZhuXF2019}, focusing \cite{LiuT, Qi, ChenJ} and cloaking \cite{Cummer2008, Zigoneanu, Norris}. More recently, the application of acoustic metamaterials to achieve nonreciprocal acoustic transmission has drawn significant attention \cite{Nassar, Rasmussen}. In a conventional acoustic medium, the transmission between two points (a source and a receiver) is unaltered upon exchanging the source and the receiver locations; in other terms the transmission of sound satisfy the principle of reciprocity \cite{Pierce}. However, this reciprocal behavior can be intentionally broken by designing nonreciprocal waveguides. These unconventional acoustic devices have been shown to hold significant potential for applications like medical imaging \cite{ZhuSL} and surface acoustic wave devices \cite{Verba, YuT}, to name a few.

To-date, several viable nonreciprocal acoustic devices have been proposed. They exploited distinct physical principles, for example, biased fluid motion \cite{Fleury2014}, spatio-temporal modulation \cite{Fleury2015, Swinteck}, strong nonlinearities \cite{Boechler, Liang2009, Liang2010}, and topological acoustics \cite{XuX, Brouzos}. Thermoacoustic (TA) coupling can also serve as another approach to break the acoustic reciprocity. Experimental demonstrations of TA diodes \cite{Biwa2016} and TA amplifiers \cite{Senga} have shown the potential for nonreciprocal propagation when using periodically distributed TA coupling elements (also known as regenerators, or REGs) in a one-dimensional (1D) waveguides. In these devices, time-harmonic waves propagating in the direction of the increasing temperature gradient (imposed by the REGs) are amplified, while the counter-propagating component is attenuated. The nonreciprocal behavior of time-harmonic waves in TA-coupled periodic waveguides has been successfully modeled and explained via a complex band structure analysis in \cite{HaoPRB}. The nonreciprocity emerging in the TAWG was also interpreted in terms of a nonreciprocal Willis material. More recently, Olivier \textit{et al.} \cite{Olivier} provided a closed-form mathematical representation of the nonreciprocity of thermoacoustic amplifiers in the framework of nonreciprocal Willis coupling.

In existing studies concerning nonreciprocal TA devices (e.g. the TA amplifier in \cite{Senga} and the infinite TA waveguide (TAWG) in \cite{HaoPRB}), time-harmonic oscillations were imposed so that every particle in the device oscillates at a constant amplitude. In these previous studies, the TA device consisted in a periodic structure whose unit cell was made up of a REG. In the complex band structure analysis \cite{HaoPRB}, the time-harmonic frequency was considered as the input variable to calculate the corresponding complex wavenumber of the infinite waveguide. The imaginary part of the wavenumber was indicative of the spatial amplification or attenuation. In the experiments reported in \cite{Senga}, alternatively, the finite TA amplifier was terminated by acoustic sensors that were tuned to match the impedance at the ends of the unit cell. Such termination should not be conflated with a non-reflecting boundary condition. Instead, it created proper reflections as if the end unit cell was connected to a neighboring unit cell, so that the characteristics of the TA waves in this finite structure were identical to those of TA Bloch waves in an infinite waveguide \cite{HaoPRB}. Such treatment assured the time harmonic character of the TA waves. However, from a practical perspective the intrinsic modes of a finite TAWG (either if terminated by a passive boundary condition, such as an anechoic termination \cite{Olivier}, a sound hard \cite{HaoJAP} or sound soft termination\cite{ChenG}, or connected to an external acoustic load \cite{Swift1998,LinJ,XuJY2022, WangK,ChenXu}), can be dynamically unstable. The unstable modes are associated with complex eigenfrequencies \cite{Guedra2012}, so that the time-harmonic response no longer takes place. This instability arises from the TA coupling elements, which also function as energy sources, actively pumping energy into the waveguide. When TA instability occurs, the amplitude of the particle oscillations in the TAWG grows with time, thus hindering its capability of sound manipulation.

The existence of TA instabilities has been known for centuries. Modern research of TA instability mainly focuses on its occurrence in (1) thermoacoustic engines \cite{Swift1998, Yazaki1998,Migliorino,LinJ,ChenG,HaoJAP}, and (2) combustion systems \cite{Keller, Hernandez, Campa}. The TA instability is beneficial for an eco-friendly energy conversion in the former case, but detrimental to the structural safety and core functions of the combustion system in the latter case. Nevertheless, the thermoacoustically unstable modes in both configurations originate either from the acoustic standing \cite{ChenG, LinJ,HaoJSV19,HaoMSSP} or traveling eigenmodes \cite{TanJQ, YuZB, XuJY} in the finite structures. In other words, the intrinsic acoustic modes exist regardless of the TA coupling, while the TA coupling elements (REGs in TA engines, or flames in combustors) destabilize them. It was only recently discovered in a study conducted on a combustion system that when a flame is inserted in a hollow waveguide terminated by sound absorbing materials (or equivalently with anechoic boundary conditions), unstable TA modes can still develop despite the large acoustic losses provided by the terminations \cite{Hoeijmakers, Emmert}. These unstable modes do not rely on any acoustic feedback provided by the structure. Recall that a straight hollow waveguide with anechoic terminations does not support acoustic eigenmodes because the right- (left-) propagating wave is prohibited by the absorbing boundary condition on the left (right) end. Given that this unstable mode was originated by the interplay between the flame and the pressure fluctuations inside the anechoically-terminated waveguide, this mode was coined as the intrinsic thermoacoustic (ITA) mode. ITA is well distinguished by the well understood unstable acoustic modes in the TA systems.

In this work, we investigate a REG-based TAWG terminated by anechoic boundary conditions. We show that the abrupt cross-sectional area change at the REG's ends gives rise to acoustic evanescent (AE) modes with either pure imaginary or complex eigenfrequency, referred to as the \enquote{overdamped} and the \enquote{oscillatory evanescent} modes, respectively. The thermoviscous effect in the REG channels leads to the occurrence of additional overdamped evanescent modes, dubbed thermoviscous evanescent (TVE) modes, which have imaginary sound speed in the REG. These TVE modes do not take place in an inviscid waveguide. With a sufficiently strong TA coupling, the local temperature gradients imposed along the REGs destabilize the AE modes, represented by a complex eigenfrequency with a negative imaginary part. The destabilized AE modes in a 1D waveguide with anechoic terminations share significant resemblance with the ITA modes observed in combustion systems, so they are referred to as the ITA modes of the TAWG. The eigenvalue analysis further shows that the onset of the TA instability is favored by either higher temperature gradients or by increasing the numbers of unit cells. Time-dependent numerical results further substantiate the existence of ITA modes.

From a general perspective, the instability associated with ITA modes is considered harmful for the performance of TAWGs (e.g. for those TAWGs designed to achieve non-reciprocal transmission). This is due to the fact that the unstable intrinsic modes may lead to large-amplitude resonances that not only threaten the structural integrity of the system, but also destroy its target functionalities. In this regard, ITA modes mitigation becomes a critical need. On the other side, the energetic nature of the ITA modes might also be leveraged for the design of a novel type of TA engines (TAEs) that could be more tunable and compact. Regardless of the specific end application, it is vital to understand the mechanism of the ITA modes, so to either conceive approaches to predicting and mitigating these modes, or to facilitate the practical implementation of the new-generation TAEs that can effectively exploit them.  This study reveals the existence and the mechanisms that control the formation of ITA modes in a finite TAWG with anechoic terminations. This configuration is especially relevant to sound propagation in finite TAWGs that are critical either for experimental validation \cite{Olivier} or for practical implementation of this concept in real-world devices. The theoretical understanding of ITA modes will also facilitate the synthesis of mitigation strategies to tame dynamic instabilities in TAWGs, thus maximizing its performance under practical operating conditions.

\section{\label{PS}Problem Statement}
\begin{figure*}
    \centering
    \includegraphics[width=\linewidth]{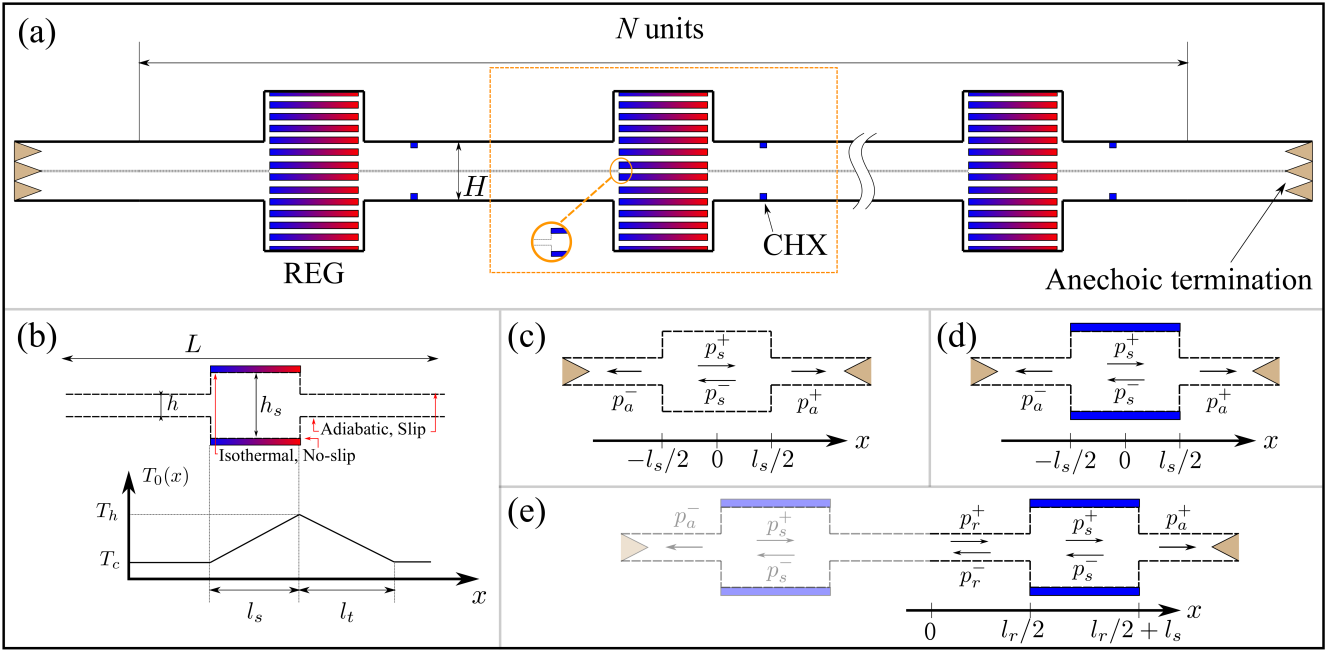}
    \caption{(a) Schematic of the $N$-unit finite TAWG with anechoic terminations. Each unit cell (outlined by the dashed box) consists of a REG that facilitates the thermoacoustic coupling. The inset in the dashed box illustrates the width change at the cold end of the REG in the minimal unit. (b) (top) The minimal unit of one unit cell in the TAWG being modeled and (bottom) the mean temperature distribution along the unit cell. A spatial temperature gradient is imposed on the REG. The forward and backward propagating components $p^+$ and $p^-$ of the pressure fields in each section of (c) an inviscid one-cell system, (d) a thermoviscous one-cell system, and (e) a thermoviscous two-cell system.}
    \label{schematic}
\end{figure*}

The benchmark system explored in this study involves a finite TAWG consisting of $N$ unit cells, as shown in Fig. \ref{schematic}(a). The TAWG is connected to a hollow waveguide and terminated by an anechoic termination on each side. Each unit cell, as outlined in the dashed box in Fig. \ref{schematic}(a), has length $L$ and consists of a regenerator (REG) of length $l_s$ located at the cell center. The REG is a porous material, which can be regarded as a stack of short parallel plates separated by thin channels. When passing through these channels, the low-frequency acoustic waves are subject to considerable thermoviscous effects. To facilitate an effective TA coupling, a spatial temperature gradient is imposed on the REG so to elevate the temperature from ambient temperature $T_c$ (blue color in REG section in Fig. \ref{schematic}(a)) at one end to the hot temperature $T_h$ (red) at the other end. The hot end of the REG is connected to a thermal buffer tube (TBT), terminated by a cold heat exchanger (CHX) that recovers the reference ambient temperature $T_c=T_\text{ref}$. Note that the TBT enables the continuous temperature distribution along the unit cell, as depicted in Fig. \ref{schematic}(b, bottom), while also acting as a local scatterer due to the temperature variation from $T_\text{ref}$.

We adopt the plane wave assumption for the wave propagating in the hollow ducts other than the REG, considering the fact that the hollow ducts are much wider than the REG channels so the thermoviscous effects near the duct walls are negligible. This assumption becomes invalid for the waves inside the REG channels due to the non-negligible thermoviscous effects. Instead, inside the REG channels, the acoustic field is described by Rott's thermoacoustic linear theory \cite{Rott,SwiftBook}:\
\begin{align}
    \frac{dp}{dx}&=-\frac{\rho_0}{1-f_v}(i\omega)u& \label{mom}\\
    \frac{du}{dx}&=-\frac{1+(\gamma-1)f_k}{\gamma P_0}(i\omega)p+g u \label{con}
\end{align}
where
\begin{equation}
    g=\frac{f_k-f_v}{(1-f_v)(1-\mathrm{Pr})}\frac{1}{T_0}\frac{dT_0}{dx} \label{g}
\end{equation}
$u$ and $p$ are first-order cross-sectionally averaged particle velocity and pressure, respectively. Note that the harmonic assumption $\mathrm{exp}(i\omega t)$ is adopted here, where $\omega=\omega_r+i\omega_i$ denotes the complex eigenfrequency, whose real and imaginary parts indicate the angular frequency and the decay rate of the acoustic oscillation, respectively. $\rho_0, P_0$ and $T_0$ are zeroth-order (mean-state) density, pressure and temperature, respectively, in the frequency domain. $\gamma$ and Pr are the specific heat ratio and the Prandtl number, respectively. $f_k$ and $f_v$ are complex thermoviscous functions expressed as: 
\begin{align}
    f_{v}&=\frac{\mathrm{tanh[(1+i)}(h_s/2)\sqrt{\omega/2\nu}]}{\mathrm{[(1+i)}(h_s/2)\sqrt{\omega/2\nu}]}\nonumber\\
    f_{k}&=\frac{\mathrm{tanh[(1+i)}(h_s/2)\sqrt{\mathrm{Pr}\omega/2\nu}]}{\mathrm{[(1+i)}(h_s/2)\sqrt{\mathrm{Pr}\omega/2\nu}]}
    \label{fvfk}
\end{align}
where $h_s$ is the width of the REG cell, and $\nu$ is the dynamic viscosity. 

When the thermoviscous effects are negligible, that is when $f_v=f_k=0$, the Helmholtz equation that leads to plane-wave solutions is recovered from Eqs.$~$\ref{mom} and \ref{con}. Considering the plane wave assumption for the wide hollow ducts (outside the REG), as well as the fact that the REG channels are identical to each other, the modeling is simplified by only calculating the acoustic field in a minimal unit, including one REG channel \cite{Gupta, HaoPRB}. The portion of the domain that is actually modeled is marked by the dashed lines in Fig. \ref{schematic}(a) and highlighted in one unit cell in Fig. \ref{schematic}(b, top). With this simplification, the width ratio $s$ of the acoustic passages is kept unchanged so that $s=h_s/h=Mh_s/H$, where $h_s, h$ and $H$ are the width of one REG channel, of the minimal unit representation of the hollow duct, and of the hollow duct, respectively (Fig. \ref{schematic}(a)). $M$ is the total number of channels in each REG. The inset in Fig. \ref{schematic}(a) shows the width change at the cold end of the REG in the minimal unit.

Before analyzing the eigenmodes in the finite TAWG, we highlight that the TAWG differs from an inviscid waveguide in two main aspects: (1) the strong thermoviscous effects in the thin REG channels, captured by the thermoviscous functions $f_v$ and $f_k$ in Eqs. (\ref{mom}) and (\ref{con}), and (2) the spatial temperature gradient imposed on the REG, captured by the unique thermoacoustic coupling term $g$ (Eq. (\ref{g})). In the following, we show that the thermoviscous effects lead to two types of evanescent modes: the acoustic evanescent (AE) modes and the thermoviscous evanescent (TVE) modes. With a sufficiently strong temperature gradient, the AE modes can be excited and eventually become dynamically unstable modes.

\section{\label{EM0dT}Evanescent modes in the TAWG with zero temperature gradient}
In this section, we first investigate a single-cell system in order to understand the thermoviscous effect of the REG on the eigenmodes. Then, we analyze a two-cell system that allows considering inter-cell interactions. In both configurations, the spatial temperature gradient is set to zero so that we can isolate the contribution of the thermoviscous effect to the eigenmodes of the TAWG. Consequently, the temperature dependent material properties become constant, that is $\nu=\mu_\mathrm{ref}/\rho_\mathrm{ref}$, and $\rho_0=\rho_\mathrm{ref}=P_0/R_\mathrm{gas}T_\mathrm{ref}$, where $\rho_\mathrm{ref}$ and $\mu_\mathrm{ref}$ are density and viscosity of air at ambient temperature $T_\mathrm{ref}$. The numerical values of $\rho_\mathrm{ref}$, $\mu_\mathrm{ref}$ and $T_\mathrm{ref}$ are given in Table \ref{para}. For a given eigenfrequency $\omega$, the thermoviscous functions $f_v$ and $f_k$ are constant along the REG channel (see Eq. (\ref{fvfk})). The thermoacoustic coupling term $gu$ in Eq. (\ref{mom}) becomes negligible due to the zero temperature gradient (see Eq. (\ref{g})).

\subsection{Evanescent modes in a one-cell system}
Considering a one-cell system with anechoic terminations, as shown in Fig. \ref{schematic} (c) or (d), the pressure fields in the range of $x<-l_s/2$ and $x>l_s/2$ are expressed as
\begin{equation}
p=\begin{cases}
    p_a^- e^{ikx}, & x<-l_s/2\\
     p_a^+ e^{-ikx}, & x>l_s/2.
  \end{cases}
\end{equation}
where $p_a^+$ and $p_a^-$ are the amplitudes of the right and left propagating components in the connecting duct (Fig. \ref{schematic}(c-e)), $k=\omega/a_0$ is the wavenumber in the hollow duct and $a_0=\sqrt{\gamma P_0/\rho_0}$ is the ambient sound speed of air.

Without losing generality, the pressure field in the REG channel $(-l_s/2<x<l_s/2)$ is developed based on Eqs. (\ref{mom}) and (\ref{con})
\begin{equation}
    p = p_s^+ e^{-ik_dx} + p_s^- e^{ik_dx}
\end{equation}
where $p_s^+$ and $p_s^-$ are the amplitudes of the right and left propagating components in the REG channel (Fig. \ref{schematic}(c-e)), $k_d=f_d k$ denotes the local wavenumber in the REG channels. $f_d$ represents the thermoviscous effect on the wavenumber, expressed as
\begin{equation}
    f_d=\sqrt{\frac{1+(\gamma-1)f_k}{1-f_v}} \label{fd}
\end{equation}
The subscript $d$ denotes the diffusion effect inside the REG channels. Clearly, $f_d$ depends on frequency via the terms $f_v$ and $f_k$ (see Eq. (\ref{fvfk})).

Applying the continuity of pressure $p$ and flow rate $U=uh$ ($u$ denotes transversely-averaged particle velocity) at $x=-l_s/2$ and $x=l_s/2$ yields:
\begin{equation}
\Big(\frac{sf_d-1}{sf_d+1}\Big)^2=\mathrm{exp}[i(2f_dkl_s)] \label{1cellk}
\end{equation}

\begin{figure}
    \centering
    \includegraphics[width=\linewidth]{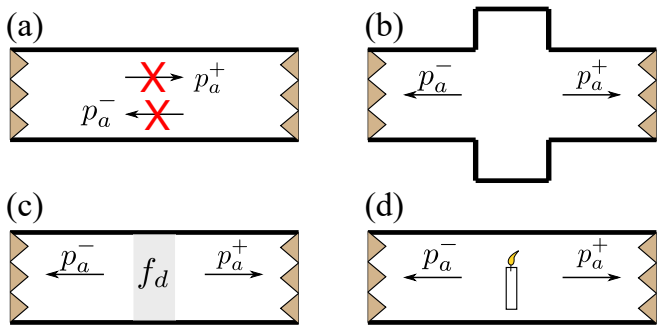}
    \caption{(a) No wave type (neither propagating nor evanescent) is allowed in a straight duct with anechoic terminations. (b-d) In the presence of heterogeneities, which can be either in the form of (b) an abrupt width change, (c) a section subject to strong thermoviscous coupling, or (d) a flame, transmitting modes become possible since the heterogeneity allows counter-propagating waves on both sides.}
    \label{ducts}
\end{figure}

The eigenfrequency $\omega$ can then be obtained from Eq. (\ref{1cellk}), considering that both $f_d$ and $k$ are dependent on $\omega$. To derive a closed-form solution, we first drop the thermoviscous effect of the REG by imposing $f_v=f_k=0$, or equivalently $f_d=1$. In practice, this aspect can be achieved by removing the short stack of plates inside the REG, so that the plane wave assumption also holds in the minimal unit of the REG section, as shown in Fig. \ref{schematic}(c). With this assumption, Eq. (\ref{1cellk}) leads to \begin{equation}
    \omega=\frac{a_0}{l_s}\Big[\mathrm{ln}\Big(\frac{s+1}{|s-1|}\Big)i+n\pi\Big] \label{inviscidom}
\end{equation}
where $n$ is an integer. The imaginary part of the eigenfrequency is always positive (that is $\omega_i=a_0/l_s\mathrm{ln[(s+1)/|s-1|]}>0$) that, according to the $\mathrm{exp}(i\omega t)$ notation, represents the evanescent modes. The time-decaying characteristics of these evanescent modes is due to the large acoustic loss provided by the two anechoic terminations. Equation (\ref{inviscidom}) reveals that in an inviscid system, the abrupt width change at the ends of the REG give rise to evanescent modes inside the waveguide with anechoic terminations. Note that Eq. (\ref{inviscidom}) does not hold if $s=1$. However, when the thermoviscous effect in the REG channels is included (i.e. $f_d\neq1$), the evanescent mode can still exist even if $s=1$ (see Eq. (\ref{1cellk})). Note that the effects of $s$ and $f_d$ on the left hand side of Eq. (\ref{1cellk}) are commutative. In either case, the appearance of the eigenmode in an anechoically-terminated waveguide relies on the presence of heterogeneity inside the structure. This heterogeneity, in the form of either an abrupt channel width change (Fig. \ref{ducts}(b)) or a local thermoviscous effect (Fig. \ref{ducts}(c)), allows the waves on both sides to be transmitted in opposite directions, so that both absorbing boundary conditions are satisfied. Looking back at the analogy with combustion systems with anechoic terminations \cite{Hoeijmakers, Emmert}, the flame serves as the local heterogeneity in the acoustic field that allows the appearance of the eigenmodes (Fig. \ref{ducts}(d)). These modes exist even if no acoustic feedback is provided by the boundaries. Remember that when this heterogeneity is absent in a straight duct, no wave modes are allowed if the duct is terminated anechoically (Fig. \ref{ducts}(a)). 

\begin{figure*}
    \centering
    \includegraphics[width=\linewidth]{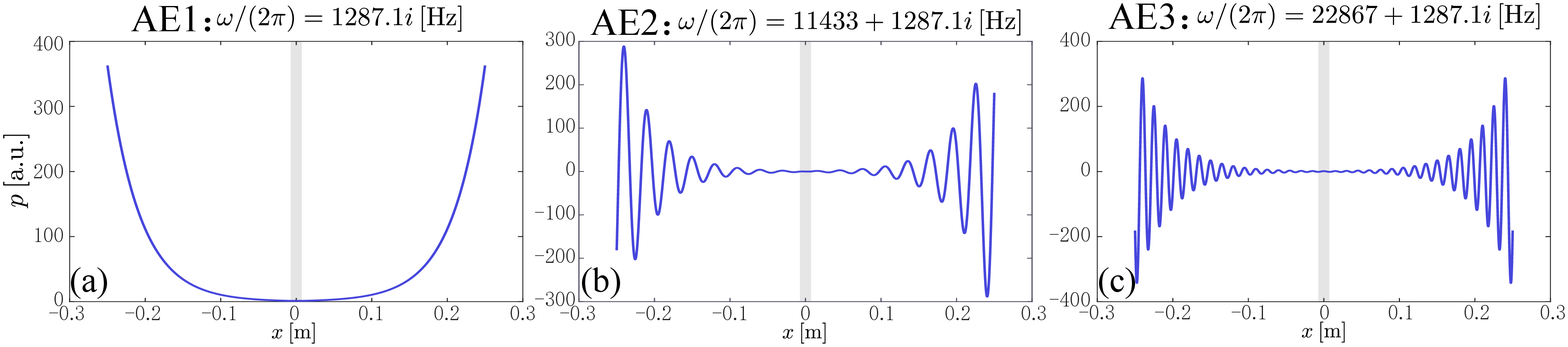}
    \caption{The pressure $p$ mode shape (in arbitrary units [a.u.]) of the lowest three acoustic evanescent (AE) modes, (a) AE1, (b) AE2, and (c) AE3, corresponding to $m=1$, $m=2$, and $m=3$, respectively, in Eq. (\ref{inviscidom}). The shaded area represents the location of the REG.}
    \label{modeshape_pa}
\end{figure*}

Based on Eq. (\ref{inviscidom}), the AE modes can be further categorized into the overdamped evanescent (OdE) mode if $n=0$, and the oscillatory evanescent (OsE) modes if $n\neq0$. The OdE mode has a pure imaginary eigenfrequency ($\omega=i\omega_i$), while the OsE modes have complex eigenfrequencies ($\omega=\omega_r+i\omega_i$). The effective wavenumber of the OsE modes is expressed as $\mathrm{Re}[k]=\mathrm{Re}[\omega]/a_0=n\pi/l_s$, which depends only on the REG length. Figure \ref{modeshape_pa} shows the OdE mode and the lowest two OsE modes ($m=$1 and 2) of the inviscid one-cell system. These three modes are labeled as AE1, AE2 and AE3, respectively. Note that the AE modes are ordered in terms of the value of $\omega_r$. The lowest AE mode (Fig. \ref{modeshape_pa}(a)) has the lowest $\omega_r$ ($\omega_r=0$ in this case), or equivalently, the longest wavelength ($\lambda=\mathrm{Re}[2\pi/k]\rightarrow\infty$). 

\begin{figure}
    \centering
    \includegraphics[width=\linewidth]{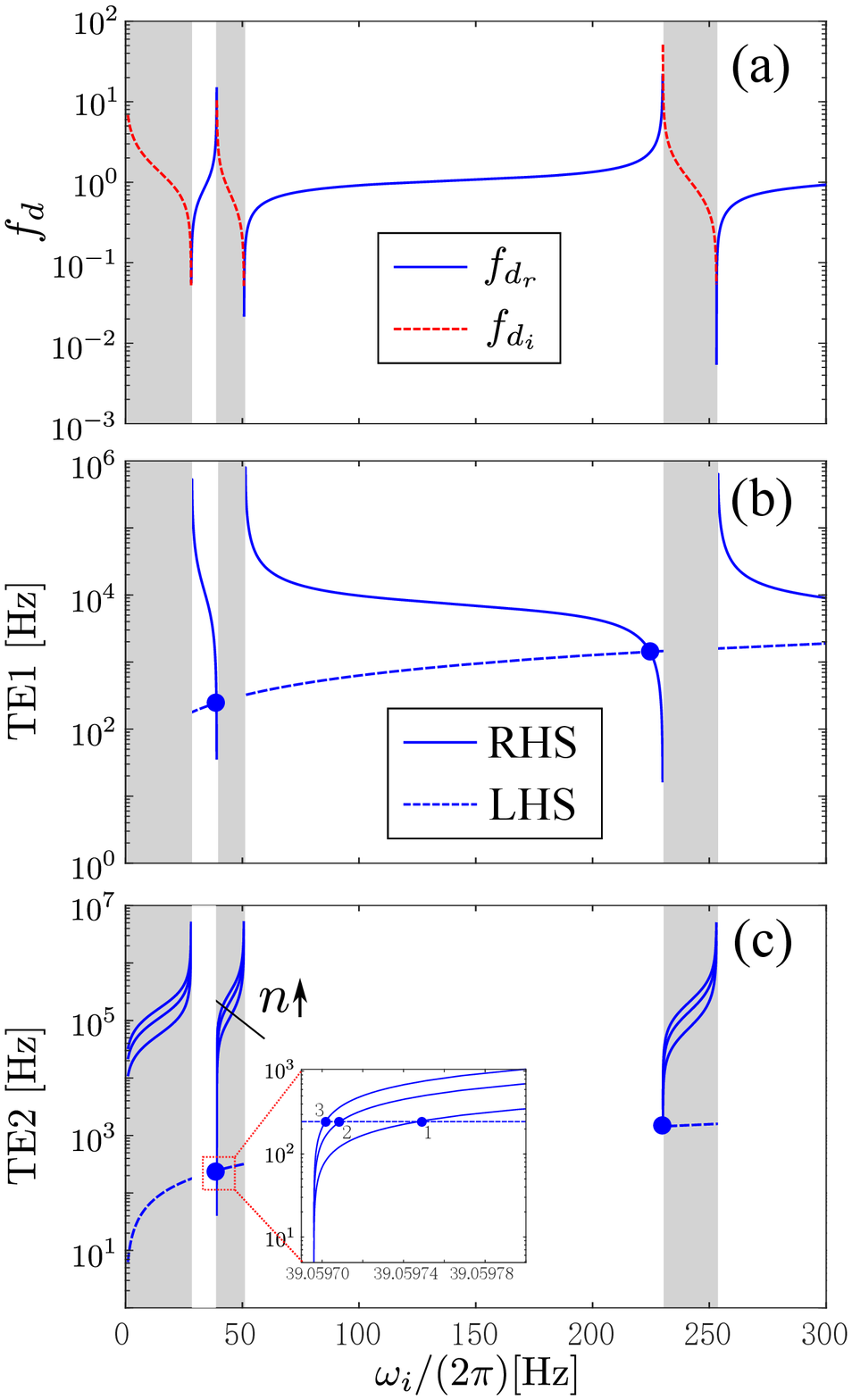}
    \caption{(a) The value of $f_d$ as a function of $\omega_i$. $f_d$ can take either purely real (blue solid) or purely imaginary (red dashed) values for a given $\omega_i$. The right hand side (RHS, solid) and left hand side (LHS, dashed) of (b) TE1 (Eq. (\ref{AEomi})) and (c) TE2 (Eq. (\ref{TVEomi})) plotted in log scale in the range of $\omega_i$ where $f_d$ is real (white area) and imaginary (gray shaded area), respectively. The inset in (c) shows the intersections of the LHS and RHS when $n$ takes the value 1, 2, and 3. The three intersections are labelled as TVE1, TVE2, and TVE3, respectively. }
    \label{OdEsol}
\end{figure}

With the initial understanding of the evanescent modes in the inviscid one-cell system, we remove the inviscid assumption and explore the thermoviscous effects ($f_d\neq1$) on the eigenmodes. We first examine the OdE modes whose eigenfrequencies are purely imaginary. Inserting $\omega=i\omega_i$ into Eq. (\ref{fvfk}) yields:
\begin{align}
    f_{v}&=\frac{\mathrm{tan}\Big[(h_s/2)\sqrt{\omega_i/\nu}\Big]}{(h_s/2)\sqrt{\omega_i/\nu}}\nonumber\\
    f_{k}&=\frac{\mathrm{tan}\Big[(h_s/2)\sqrt{\mathrm{Pr}\omega_i/\nu}\Big]}{(h_s/2)\sqrt{\mathrm{Pr}\omega_i/2\nu}}
    \label{fvfk_OdE}
\end{align}
Equation (\ref{fvfk_OdE}) suggests that the thermoviscous functions $f_v$ and $f_k$ become purely real with an imaginary $\omega$. As a result, $f_d$ can be either real ($f_d=f_{d_r}$) or imaginary ($f_d=if_{d_i}$) depending on the sign of the expression under the root square in Eq. (\ref{fd}). Fig. \ref{OdEsol}(a) shows the value of $f_d$ as a function of $\omega_i$. The numerical values of relevant parameters used for the calculation are listed in Table \ref{para}. 
\begin{table}[h!]
    \centering
        \begin{tabular}{>{\centering}p{0.1\textwidth}>{\centering}p{0.1\textwidth}>{\centering}p{0.07\textwidth}>{\centering}p{0.07\textwidth}<{\centering}p{0.07\textwidth}}
        \Xhline{3\arrayrulewidth}
        \\[-1em]
         $L\mathrm{[m]}$&$h_s\mathrm{[mm]}$ &$s$&$l_s \mathrm{[m]}$&$P_0\mathrm{[Pa]}$\\
         0.5&0.96&5.714&0.015& 101325 
         \\[-1em]
         \\\Xhline{3\arrayrulewidth} 
         \\[-1em]
         $\mu_\mathrm{ref}\mathrm{[Pa\cdot s]}$&$T_\mathrm{ref}\mathrm{[K]}$  &{$\rho_\mathrm{ref}\mathrm{[kg/m^3]}$}& Pr &$\gamma$\\
         $1.98\times 10^{-5}$ & 300 & 1.2 & 0.72 & 1.4\\
         \\[-1em]
         \Xhline{3\arrayrulewidth} 
    \end{tabular}
    \caption{Geometrical and material parameters of the thermoviscous one-cell system.}
    \label{para}
\end{table}

When $f_d$ is real ($f_d=f_{d_r}$), Eq. (\ref{1cellk}) is recast as:
\begin{equation}
\text{TE1: }\omega_i=\frac{a_0}{l_sf_{d_r}}\mathrm{ln}\Big(\Big\rvert\frac{sf_{d_r}+1}{sf_{d_r}-1}\Big\rvert\Big) \label{AEomi}
\end{equation}
TE1 stands for transcendental equation 1; both the left and the right hand terms of the equation are plotted in Fig. \ref{OdEsol}(b) versus frequency $\omega/2\pi$ (or, effectively, the unknown $\omega$) to seek a graphical solution to Eq. (\ref{AEomi}). Interestingly, the OdE mode of the inviscid one-cell system (described by Eq. (\ref{inviscidom}) ($n=0$)) can be recovered from Eq. (\ref{AEomi}) by eliminating the thermoviscous contribution, or equivalently, by imposing $f_d=f_{d_r}=1$. However, the OsE modes of the thermoviscous one-cell system, associated with complex eigenfrequencies, cannot be directly obtained from Eq. (\ref{inviscidom}). Discussions about the OsE modes will be developed later in this section. 

When $f_d$ is imaginary, or $f_d=if_{d_i}$, Eq. (\ref{1cellk}) gives rise to
\begin{equation}
\text{TE2: }\omega_i=\frac{a_0}{l_sf_{d_i}}\Big[2\mathrm{tan}^{-1}(sf_{d_i})+(n-1)\pi\Big], \quad n=0,1,2,... \label{TVEomi}
\end{equation}
Remarkably, Eq. (\ref{TVEomi}) leads to another set of solutions to Eq. (\ref{1cellk}), which is unique to the thermoviscous configuration. Therefore, the OdE modes described specifically by Eq. (\ref{TVEomi}) are dubbed thermoviscous evanescent (TVE) modes. Notably, the local wavenumber in the REG of the TVE modes turns to a real quantity, i.e., $k_d=f_dk=-f_{d_i}\omega_i/a_0$ despite that the eigenfrequency is imaginary. In this sense, the local sound speed in the REG, $c_d=\omega/k_d$ becomes an imaginary quantity, indicative of the dissipation of motion in time (yet no attenuation in space, reflected by a real $k_d$).

Figures \ref{OdEsol}(b) and (c) show the left hand side (LHS) and right hand side (RHS) of Eqs. (\ref{AEomi}) and (\ref{TVEomi}), respectively. The white (gray) area in Fig. \ref{OdEsol} indicates the frequency range where $f_d$ is real (imaginary), or equivalently, where Eq. (\ref{AEomi}) (Eq. (\ref{TVEomi})) holds true. The solutions to Eqs. (\ref{AEomi}) and (\ref{TVEomi}) are obtained by a graphical method and shown as the blue dots in Figs. \ref{OdEsol}(b) and (c), respectively. Due to the periodic nature of the function $f_d$ (the tangent function is periodic, see Eq. (\ref{fvfk_OdE})), both equations have infinite numbers of solutions. For practical considerations, we focus on the evanescent modes that have the smallest $\omega_i$, which represent the least stable modes because they are more likely to be excited by thermoacoustic instabilities. Also interesting is the fact that the solutions to Eq. (\ref{TVEomi}) with distinct values of $n$ are extremely close to each other. The inset in Fig. \ref{OdEsol}(c) shows the zoom-in details near the intersections of the solid curves (RHS of Eq. (\ref{TVEomi}) with $n$ taking the value 1, 2 and 3) with the dashed curve (LHS of Eq. (\ref{TVEomi})). The inset clearly suggests that the (imaginary) eigenfrequency of these modes are indistinguishable up to a four-digit precision. 

\begin{figure*}
    \centering
    \includegraphics[width=\linewidth]{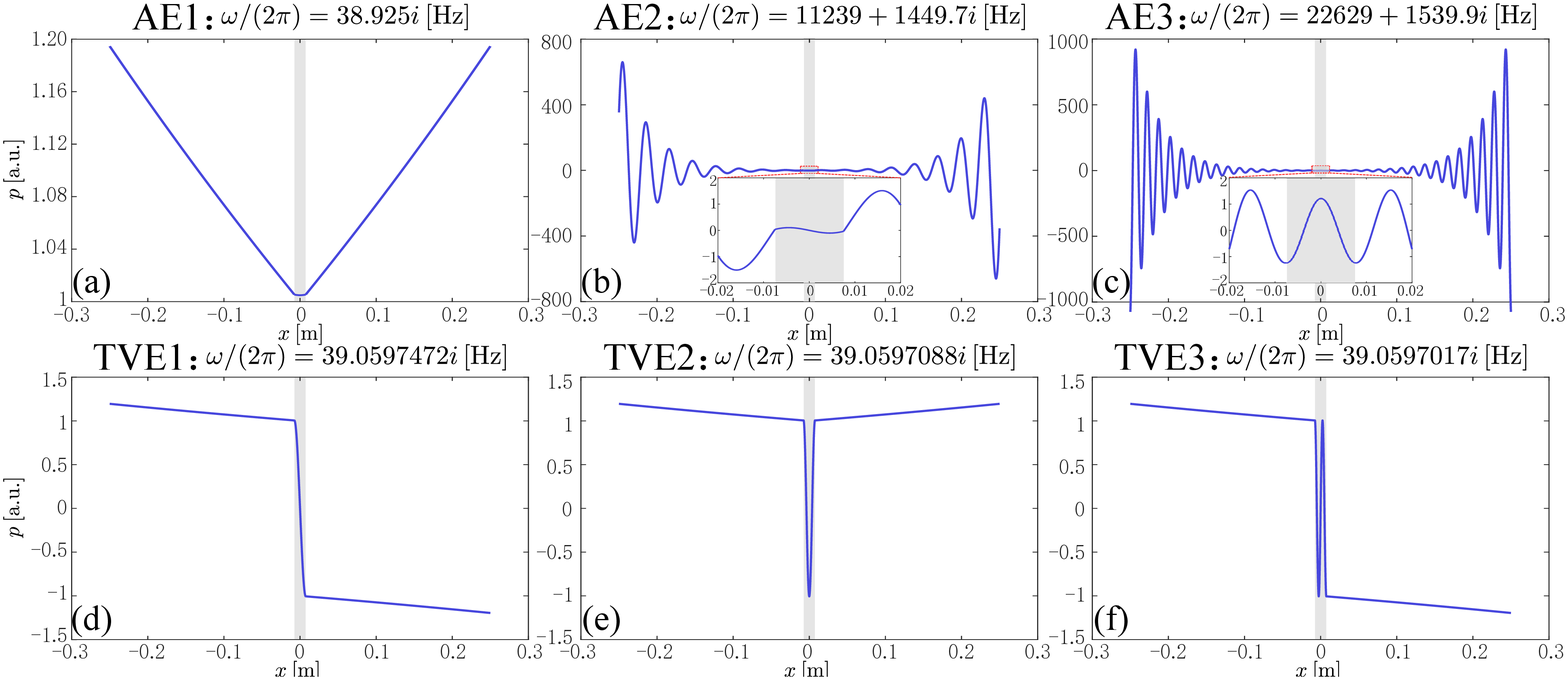}
    \caption{The mode shapes of (a-c) the lowest three acoustic evanescent (AE) modes, and (d-e) the lowest three thermoviscous evanescent (TVE) modes. The eigenfrequencies $\omega$ of the OdE modes AE1, TVE1, TVE2, and TVE3 are graphically calculated from Fig. \ref{OdEsol}(b) and (c). The eigenfrequencies of the OsE modes, AE2 and AE3, are calculated by numerically solving Eq. (\ref{log}).}
    \label{modeshape_tv}
\end{figure*}

The aggregation of these modes is due to the fact that the solutions appear near the boundary of the gray area, which is also where the value of $f_d$ approaches its singularity (Fig. \ref{OdEsol}(a)). As a result, a little perturbation of $\omega_i$ can lead to a drastic change of the value of $f_d$. Near the singularity, $f_d\rightarrow \infty$, so the LHS of Eq. (\ref{1cellk}), $[(sf_d-1)/(sf_d+1)]^2\approx 1$. Considering that $\omega=\tilde{\omega}_i i$ is a solution to Eq. (\ref{1cellk}), there exists a small perturbation $\delta_\omega$ so that $[2f_d(\tilde{\omega_i}+\delta_\omega)k(\tilde{\omega_i}+\delta_\omega)l_s]=[2f_d(\tilde{\omega_i})k(\tilde{\omega_i})l_s+2m\pi]$, where $m$ is a non-zero integer. With the perturbed value of $\omega_i$, Eq. (\ref{1cellk}) still holds with the value of its LHS being almost unchanged (approximately equal to 1). Consequently, despite the hardly distinguishable eigenfrequencies of the TVE modes, their corresponding $f_d$ functions (or, equivalently, local wavenumbers $k_d=f_d\omega/a_0$) can differ considerably. Figures \ref{modeshape_tv} (d-e) show the mode shapes of the lowest three TVE modes, labelled as TVE1, TVE2 and TVE3, respectively in Fig. \ref{OdEsol}(c). The difference in the local wavelength $k_d$ of the three modes is clearly demonstrated. The TVE modes are ordered in terms of the local wavelength of the REG, $\lambda_d=2\pi/k_d$, so that the lowest TVE mode has the longest $\lambda_d$. Remember $k_d$ is real for TVE modes, although $\omega$ is imaginary. As shown in Fig. \ref{modeshape_tv} (d-e), the local wavelength $\lambda_d$ of the REG is $4l_s$, $2l_s$, and $4/3l_s$ for the TVE1, TVE2, and TVE3 modes, respectively.

Note that solutions to Eq. (\ref{TVEomi}) are not found when $n=0$. According to the definition in Eq. (\ref{fd}), $f_{d_i}>0$ when $f_d$ is an imaginary function. As a result, the RHS of Eq. (\ref{TVEomi}) will be negative when $n=0$, since $0<\mathrm{tan}^{-1}(sf_{d_i})<\pi/2$, thus not intersecting with the positive LHS, $\omega_i$.

To obtain the OsE modes of the thermoviscous one-cell system, we rewrite Eq. (\ref{1cellk}) as:
\begin{equation}
    \mathrm{Ln}\Big[\Big(\frac{sf_d-1}{sf_d+1}\Big)^2\Big]=i(2f_dkl_s-2m\pi), \quad m =1,2,3,... \label{log}
\end{equation}
where $\mathrm{Ln}(z)$ indicates the principal value of the natural logarithm of the complex quantity $z$. Equation (\ref{log}) can be solved numerically with a nonlinear root-finding package (we used the \textit{fsolve} package in MATLAB). Note that the $m=0$ case is not included in Eq. (\ref{log}), since it leads to the solutions of either Eq. (\ref{AEomi}) or Eq. (\ref{TVEomi}). 

Figure \ref{modeshape_tv}(b) and (c) show the eigenfunctions (mode shapes) of the OsE modes associated with the complex eigenfrequencies, calculated from Eq. (\ref{log}) when $m=1$ and $m=2$, respectively. The mode shapes plotted in Fig. \ref{modeshape_tv} (a-c) show an evident resemblance with their counterparts of the inviscid one-cell system, plotted in Fig. \ref{modeshape_pa}. Hence, the modes plotted in Fig. \ref{modeshape_tv} (a-c) are referred to as the acoustic evanescent (AE) modes of the thermoviscous one-cell system: AE1, AE2, and AE3, respectively. Remember that the AE modes exist in the waveguide whether the thermoviscous effect takes place or not, while the TVE modes do not appear in an inviscid system. Figure \ref{modeshape_tv} further suggests that the mode shapes are either even or odd, in terms of their spatial symmetry. For example, AE1, AE3, and TVE2 are even, while AE2, TVE1, and TVE3 are odd. In fact, the anechoic terminations on both sides require that the specific impedance $p/u$ in the duct on the right (left) hand side of the REG is equal to $z_0$ ($-z_0$), where $z_0$ is the characteristic acoustic impedance of air. Hence, the specific impedance along the one-cell waveguide has to be an odd function of the spatial coordinate $x$, achieved by either an oddly distributed $p$ with an evenly distributed $u$, or an evenly distributed $p$ with an oddly distributed $u$.

\subsection{Evanescent modes in a two-cell system}
Following the analytical assessment of the evanescent modes in a one-cell system, we investigate the modes in a two-cell system, as sketched in Fig. \ref{schematic}(e), to understand the effect of inter-cell interactions on the eigenmodes.

Considering the symmetry of the two-cell system, the pressure field in the $x>0$ range (see Fig. \ref{schematic}(e)) is expressed as:
\begin{equation}
p=\begin{cases}
    p_r(e^{-ikx}\pm e^{ikx}), & 0<x<l_r/2\\
    p_s^+ e^{-ik_dx} + p_s^- e^{ik_dx}, & l_r/2<x<l_r/2+l_s\\
     p_a^+ e^{-ikx}, & x>l_r/2+l_s.
  \end{cases}
\end{equation}
where the $\pm$ sign is replaced by a $+$ sign considering $p$ is an even distribution about $x=0$, or equivalently, $dp/dx|_{x=0}=0$. Instead, the $-$ sign holds if $p$ is an odd distribution about $x=0$, or equivalently, $p|_{x=0}=0$. The continuity constraints of pressure $p$ and flow rates $U$ at $x=l_r/2$ and $x=l_r/2+l_s$ give rise to:
\begin{widetext}
\begin{equation}
    \Big(\frac{sf_d-1}{sf_d+1}\Big)\Bigg(\frac{\pm(sf_d-1)\mathrm{exp}[ik(l_r/2)]+(sf_d+1)\mathrm{exp}[-ik(l_r/2)]}{\pm(sf_d+1)\mathrm{exp}[ik(l_r/2)]+(sf_d-1)\mathrm{exp}[-ik(l_r/2)]}\Bigg)=\mathrm{exp}[i(2f_dkl_s)] \label{2cellk}
\end{equation}
\end{widetext}
According to Fig. \ref{schematic}(e) and Table \ref{para}, the length of the connecting duct $l_r=L-l_s=0.485\:\mathrm{[m]}$. Equation (\ref{2cellk}) is then solved graphically for the OdE modes, and numerically with a nonlinear root-finding package for the OsE modes, respectively. 

\begin{figure*}
    \centering
    \includegraphics[width=\linewidth]{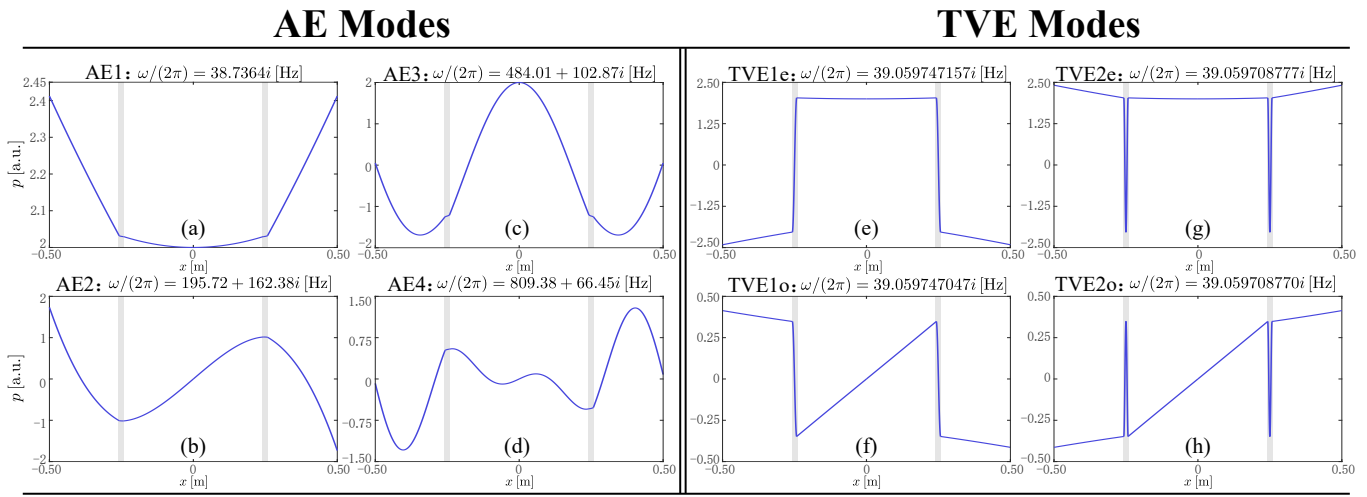}
    \caption{The eigenfrequencies and mode shapes of (a-d) the lowest four AE modes, and (e-h) the lowest four TVE modes of the two-cell system. The top and bottom rows plot the modes that are even and odd distributions of $p$, respectively.}
    \label{modeshape_2}
\end{figure*}

Figure \ref{modeshape_2} summarizes the eigenfrequencies and eigenmodes of the lowest four AE modes (Figs. \ref{modeshape_2} (a-d)) and the lowest four TVE modes (Figs. \ref{modeshape_2} (e-h)). The top and bottom rows show the even and odd symmetric modes of pressure $p$, respectively. For a waveguide that consists of even number of unit cells, e.g., $N=2$, the TVE modes appear in pairs for a given $\lambda_d$. For example, Figs. \ref{modeshape_2} (e) and (f) plot the even and the odd modes that have identical $\lambda_d=4l_s$, thus labeled as TVE1e and TVE1o modes, respectively. The eigenfrequencies of these two modes are hardly distinguishable  for practical considerations, and approximately equivalent to the eigenfrequency of the TVE1 mode of the one-cell system (see Fig. \ref{modeshape_tv}(d)). By comparing Fig. \ref{modeshape_tv} and Fig. \ref{modeshape_2}, the eigenfrequencies of the TVE modes are not altered significantly by the inter-cell interaction, in that they are primarily determined by the local wavelength $\lambda_d$ of the thermoviscous channel.
However, the eigenfrequencies of the AE modes of the two-cell system are considerably different from their counterparts of the one-cell system. Comparing Figs. \ref{modeshape_2} (a-d) with Figs. \ref{modeshape_tv}(a-c), a significant difference in the frequency $\omega$ and the wavelength $\lambda$ of the OsE modes is observed. In the two-cell system, the two thermoviscous channels are connected by an inviscid duct (the section $-l_s/2<x<l_s/2$ shown in Fig. \ref{schematic}(e)), so that they construct a resonating structure in which the forward and backward propagating waves can interfere constructively to form a (time-decaying) standing wave pattern. Figure \ref{modeshape_2}(b) shows the mode shape of the lowest oscillatory AE mode, or AE2. Remember that AE1 is an OdE mode, shown in Fig. \ref{modeshape_tv}(a). The wavelength of the AE2 mode is $\lambda=4(l_r+2l_s)$, that is much larger than its counterpart in the one-cell system ($\lambda=4l_s$, see Fig. \ref{modeshape_tv}(b)). As a result, the angular frequency, $\mathrm{Re}[\omega]$ corresponding to this mode is effectively reduced in the two-cell system. In fact, with more unit cells connected to construct the waveguide, the resonating structure becomes longer, which leads to the reduction of the angular frequency in each mode that is higher than AE1.

\section{\label{ITAmodes}Intrinsic thermoacoustic modes in the TAWG with a non-zero temperature gradient}

In this section, we will explore the evolution of the evanescent modes with an increasing temperature gradient applied on the REG. The spatial temperature gradient, accompanied by the thermoviscous effect in the REG channels, constructs the complete thermoacoustic coupling, represented by the $gu$ term in Eq. (\ref{con}). 

With a spatially varying temperature distribution $T_0$, the thermoviscous functions $f_v$ and $f_k$ are no longer constant along the REG channel for a given $\omega$, since the dynamic viscosity $\nu$ is temperature dependent (see Eq. (\ref{fvfk})). Therefore, Eqs. (\ref{mom}) and (\ref{con}) do not allow analytical solutions. Instead, we conduct the analyses via the finite element method (FEM) using the commercial software COMSOL Multiphysics. In the FEM model, we numerically solve the 2D conservation equations of continuity, momentum and energy for the REG channels, from which the acoustic pressure $p$, the particle velocity field $\mathbf{u}$, and the perturbation temperature $T$ can be obtained. The boundaries of the REG channels are defined as no-slip, isothermal walls (Fig. \ref{schematic}(b, top)) to enforce the thermoviscous diffusion in the channels. The acoustic pressure field in the inviscid ducts that connect the REG channels is obtained by solving the 2D Helmholtz equation. The temperature profile $T_0$ (Fig. \ref{schematic}(b, bottom)) is assigned as the base state to each unit cell in the TAWG. The normal impedance at the two ends of the TAWG is set to $z_0$, so that no reflection is allowed at the boundaries.

\subsection{Eigenvalue analyses of the TAWG imposed with a non-zero temperature gradient}
\begin{figure*}
    \centering
    \includegraphics[width=\linewidth]{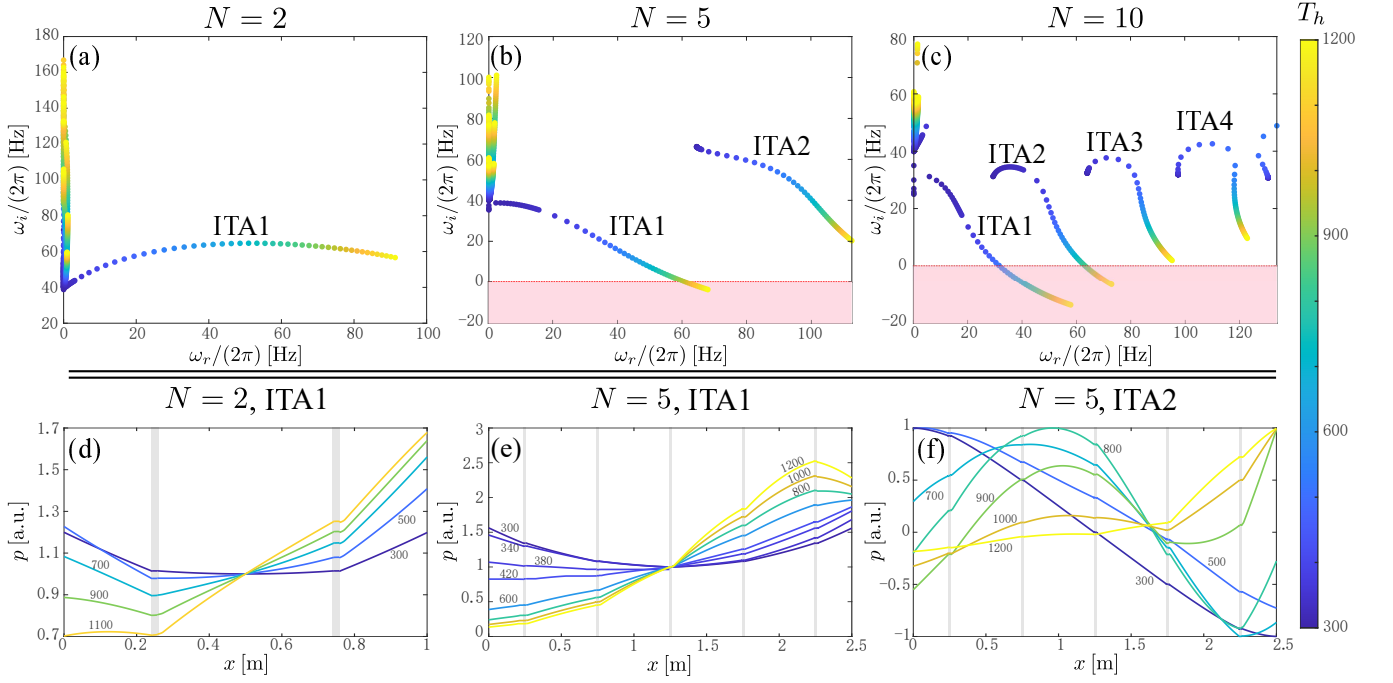}
    \caption{The eigenfrequency plots of the (a) two-cell ($N=2$), (b) five-cell ($N=5$), and (c) ten-cell ($N=10$) TAWGs under varying hot temperature $T_h$ imposed on the REGs. In (a-c), the horizontal and vertical axes denote the real and imaginary parts of the eigenfrequency $\omega/(2\pi)=\omega_r/(2\pi)+i\omega_i/(2\pi)$, respectively. The pink shaded areas in (b) and (c) denote the unstable regions, where $\omega_i<0$. The mode shapes of (d) the ITA1 mode of the two-cell system, (e) the ITA1 mode of the five-cell system, and (f) the ITA2 mode of the five-cell system, at selected values of $T_h$. The gray shaded areas in (d-f) indicate the locations of the REGs in each system. The mode shapes plotted in (d) and (e) are normalized so that the pressure at the mid-point is $p=1\:\mathrm{[a.u.]}$. The mode shapes in (f) are normalized by the maximal absolute value $|p|$ of each curve for easier visual comparison. The dots in (a-c) and the curves in (d-f) are colored by the value of $T_h$ (see the color bar). }
    \label{eigfreq}
\end{figure*}

To calculate the complex eigenfrequencies of the $N$-cell TAWG, we conduct eigenvalue analyses, in which the hot-end temperature $T_h$ of each cell (Fig. \ref{schematic}(b)) was input as a variable that represents the strength of the TA coupling. Figures \ref{eigfreq}(a-c) show the complex eigenfrequencies $\omega/(2\pi)=\omega_r/(2\pi)+i\omega_i/(2\pi)$ of a two-cell ($N=2$), five-cell ($N=5$), and ten-cell ($N=10$) TAWG, respectively. The horizontal and vertical axes represent the real and imaginary parts of $\omega/(2\pi)$, respectively. $T_h$ was continuously varied from 300 [K] to 1200 [K] to generate Figures \ref{eigfreq}(a-c). In all three cases, the modes with angular frequency in the selected range $\omega_r/(2\pi)<140\:\mathrm{[Hz]}$ are plotted.

In each case shown in Fig. \ref{eigfreq}(a-c), a cluster of solutions appears near (or on) the imaginary axis, despite of the variation of $T_h$. The eigenfrequencies of these modes are dominated by their positive imaginary parts. By inspecting their mode shapes (not shown in Fig. \ref{eigfreq}), we observed that these modes evolve from the TVE modes analyzed in Section \ref{EM0dT}. Remember when $T_h=T_\mathrm{ref}=300\:\mathrm{[K]}$, the TAWG is equivalent to the thermoviscous waveguide that were investigated in Section \ref{EM0dT}. When subject to a stronger TA coupling (higher $T_h$), these evanescent modes become more stable, which is represented by an increasing $\omega_i$. Therefore, they are practically trivial, since the time-decaying nature of these modes are not altered by the TA coupling.

However, with the appearance of a non-negligible TA coupling, a unique type of solutions appears that originates from the AE modes. In the two-cell TAWG ($N=2$, Fig. \ref{eigfreq}(a)), the mode labeled with ITA1 that has a purely imaginary eigenfrequency when $T_h=300\:\mathrm{[K]}$ acquires an oscillatory motion in time, represented by a non-zero $\omega_r$; this behavior becomes more pronounced as $T_h$ increases. By checking the evolution of the mode shape with $T_h$, shown in Fig. \ref{eigfreq}(d), we realize that this mode is the AE1 mode of the two-cell system, when $T_h=300\:\mathrm{[K]}$.  For comparison purposes, we normalized all the mode shapes in Fig. \ref{eigfreq}(d), so that the value of the pressure eigenfunction at the midpoint of the TAWG is $p=1\:\mathrm{[a.u.]}$, where a.u. denotes arbitrary units. These modes that appear due to the TA coupling are therefore referred to as the intrinsic thermoacoustic, or ITA, modes. When the TA coupling is activated, the even symmetry of AE1 is broken, but a clear connection between the mode shape of the AE1 mode (Fig. \ref{modeshape_2}(a)) and that of the ITA1 modes shown in Fig. \ref{eigfreq}(d) is observed. Specifically, the ITA1 mode at $T_h=300\:[\mathrm{K}]$ shown in Fig. \ref{eigfreq}(d) is identical to the AE1 mode plotted in Fig. \ref{modeshape_2}(a). As $T_h$ is increased, the mode shape is gradually deformed as the even symmetry is broken (Fig. \ref{eigfreq}(d)). We will show later that each AE mode in the $N$-cell thermoviscous system will evolve towards an ITA mode when the TA coupling is activated, or equivalently $T_h\neq T_c$. For the two-cell system, the angular frequency $\omega_r$ of the AE2 mode is 195.72 [Hz] (Fig. \ref{modeshape_tv}(c)), so the higher AE modes, as well as their associated ITA modes, do not appear in the displayed range of frequency in Fig. \ref{eigfreq}(a).

Now, we consider a TAWG consisting of five unit cells ($N=5$). Figure \ref{eigfreq}(b) plots the eigenfrequencies of this five-cell system. Similar to the two-cell system, the ITA1 mode (that is also the AE1 mode when $T_h=300\:\mathrm{[K]}$) appears. Figure \ref{eigfreq}(e) shows the mode shapes of the ITA1 mode at selected $T_h$ values. At $T_h=300\:\mathrm{[K]}$, the eigenfunction $p$ of the ITA1 mode, or equivalently the AE1 mode, is even-symmetric. With the increase of $T_h$, the even symmetry is broken, as also seen in the two-cell system (Fig. \ref{eigfreq}(d)). In Fig. \ref{eigfreq}(b), an additional curve appears. By tracking the evolution of the eigenfunction of this mode with $T_h$, shown in Fig. \ref{eigfreq}(f), we confirm that this mode is the ITA2 mode. Remember that the ITA2 mode is identical to the AE2 mode when $T_h=300\:\mathrm{[K]}$. Note the odd symmetry of the eigenfunction $p$ when $T_h=300\:\mathrm{[K]}$.
In a thermoviscous waveguide, the angular frequency $\omega_r$ of AE modes (that are higher than AE1) decreases with the increasing number $N$ of unit cells, due to the increased length of the resonating structure, as concluded in Section \ref{EM0dT}. This aspect explains why in the displayed angular frequency range, only one ITA mode appears in the two-cell TAWG, but two ITA modes appear in the five-cell configuration. Analogously, increasing the cell number to $N=10$ lets four ITA modes appear in the angular frequency range of $\omega_r/(2\pi)<140\:\mathrm{[Hz]}$, as shown in Fig. \ref{eigfreq}(c).

In the five-cell TAWG, as the TA coupling is sufficiently strong, the decay rate $\omega_i$ of the ITA1 mode turns to negative value (Fig. \ref{eigfreq}(b)), which is indicative of the onset of TA instability. The onset temperature $T_h$ of the dynamic instability of the ITA1 mode is approximately $T_{h_\mathrm{on}}=980\:\mathrm{[K]}$. Remember the onset temperature of TA instability is the hot temperature at which the TA system is marginally stable \cite{Guedra,TanJQ}. With a hot end temperature $T_h>T_{h_\mathrm{on}}$, the ITA1 mode is unstable. 
In the ten-cell system, as shown in Fig. \ref{eigfreq}(c), the onset temperature of the ITA1 mode is significantly reduced to about $T_{h_\mathrm{on}}=520\:\mathrm{[K]}$. Remarkably, in the ten-cell system, the ITA2 mode also becomes unstable at $T_{h_\mathrm{on}}=820\:\mathrm{[K]}$. In theory, the two-cell system may also sustain unstable ITA modes granted that the TA coupling is sufficiently strong. We tracked the evolution of the eigenfrequency of the ITA1 mode of the two-cell system with varying $T_h$ until $T_h=10000\:\mathrm{[K]}$, which is far beyond the temperature range displayed in Fig. \ref{eigfreq}(a). At $T_h=10000\:\mathrm{[K]}$, $\omega/(2\pi)=163.63+6.876i\:\mathrm{[Hz]}$. Compared to the eigenfrequencies plotted in Fig. \ref{eigfreq}(a), the decay rate $\omega_i$ is much closer to the boundary of the unstable region, i.e. $\omega_i=0$. Nevertheless, this range is of low practical relevance given that this temperature exceeds the operating range of most solid materials typically used to build the REG.

Based on the above observations, we draw the conclusion that the tendency of a TAWG to become dynamically unstable increases either (1) when in presence of high hot end temperatures $T_h$, or (2) with an increasing number of unit cells. The REGs in the TAWG are TA coupling elements that are capable of injecting thermal energy into the sound field. On one hand, a higher $T_h$, or effectively a steeper spatial temperature gradient, increases the potential for thermo-acoustic energy conversion. On the other hand, the increased number of unit cells brings an increased number of energy sources (that is of REGs), thus favoring the onset of the TA instability. We merely note that, in the TA community, it is a common practice to construct a multiple REG, or multi-stage, system in order to enhance the performance of TA devices \cite{XuJY,XuJY2022,WangK,HaoJAP}. Despite the fact that the TA instability is a key mechanism for TA energy conversion devices, it is detrimental to the application of TAWGs. Considering the periodic nature of TAWGs, involving a large number of unit cells is inevitable in the design and construction of TAWGs. Therefore, specific measures to mitigate TA instabilities have to be taken into account in the design process of TAWGs. We note that extensive studies on the suppression of TA instability have been conducted in the combustion literature. Several effective instability suppression measures have been proposed, including the use of passive (e.g., acoustic liners \cite{Kelsall} or Helmholtz resonators \cite{ZhaoD}) or active \cite{McManus} elements. Although the design of instability suppression in the TAWGs is beyond the scope of this study, the author believes that the existent TA noise mitigation techniques that are effectively applied to combustion systems are applicable to the design of TAWGs.

\subsection{Time-dependent simulations of the unstable ITA modes}
\begin{figure*}
    \centering
    \includegraphics[width=\linewidth]{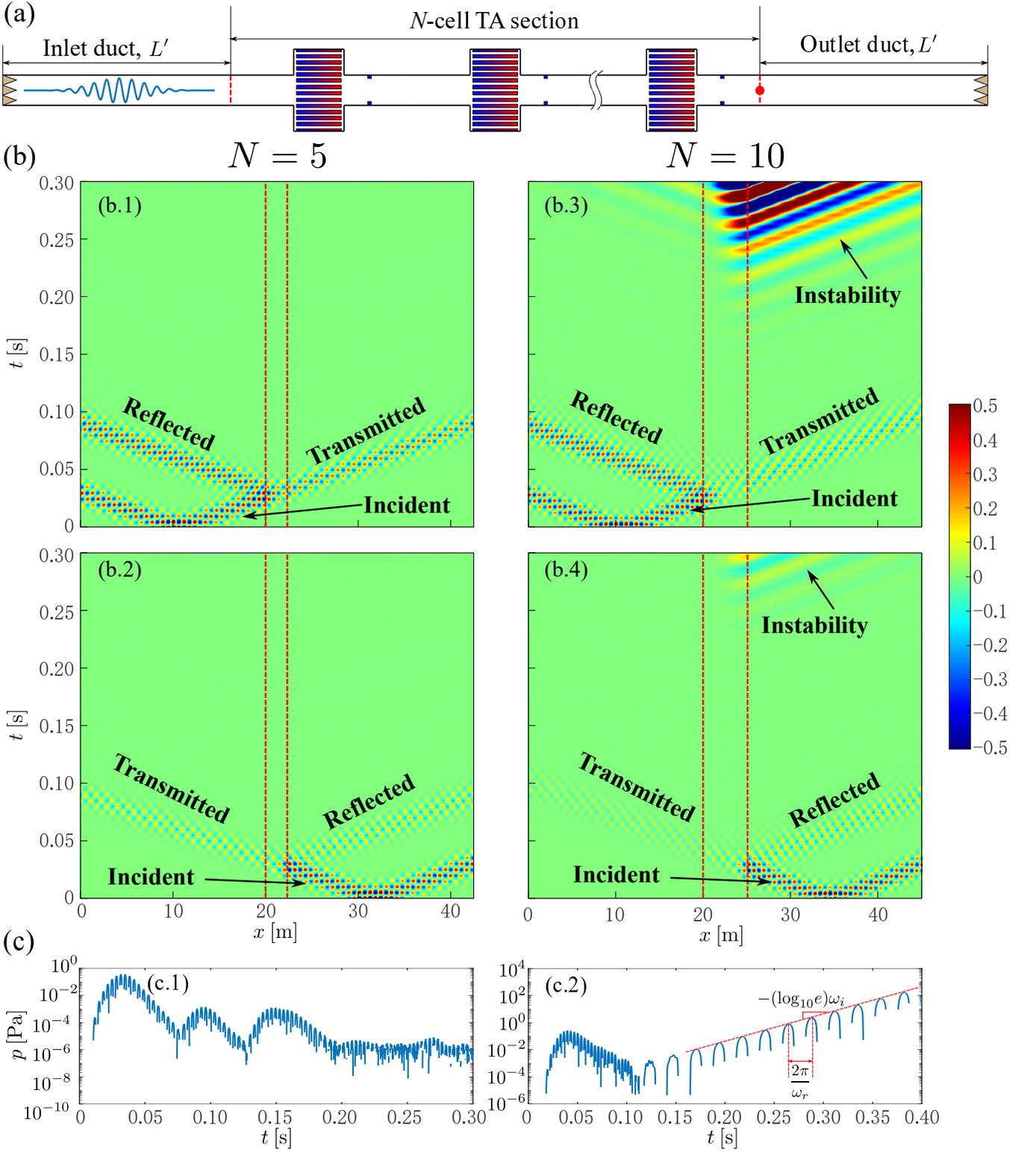}
    \caption{(a) The schematic of an $N$-cell TA section connected to inlet and outlet ducts terminated by anechoic boundary conditions. At $t=0$, a wave packet is released in the inlet duct. Only the case where the wave packet impinges from the left hand side of the TA section is shown. (b) The space-time plot of the five-cell ($N=5$)/ten-cell ($N=10$) TAWG in which the wave packet impinges from the (b.1)/(b.3) left and (b.2)/(b.4) right hand side of the TA section, respectively. The hot temperature $T_h$ imposed on the REGs is $T_h=750\:\mathrm{[K]}$. (c) The time response of the acoustic pressure evaluated at the intersection of the TA section and the outlet duct (marked as the red dot in (a)) in the case that the wave packet impinges from the left hand side of the (c.1) five-cell and (c.2) ten-cell TA section. The acoustic pressure is plotted in log scale, so the negative-valued pressure is neglected.}
    \label{time}
\end{figure*}

To corroborate the TA instability of the ITA modes predicted by the eigenvalue analyses, we will evaluate the time response of the TAWG under an incident wave packet. In the time-dependent simulations, we attach an inlet duct and an outlet duct of length $L^\prime=20\:\mathrm{[m]}$ to each side of the $N$-cell TA system, shown in Fig. \ref{time}(a). The two ducts are terminated by anechoic boundary conditions. At time instant $t=0$, a spatial wave packet, shown in Fig. \ref{time}(a) is enforced as an initial condition. The windowed wave packet $p_{wp}$ is expressed as:
\begin{equation}
    p_{wp}=w_H(x-\xi)p_g(x-\xi)
\end{equation}
where $\xi$ is the x-coordinate of the center of the windowed pulse, $w_H$ is a Hann window expressed as: $w_H=(\mathrm{cos}[2(x-\xi)\pi/D]+1)/2$, $D$ is the total length of the Gaussian pulse $p_g$. $p_g$ has a carrier wavenumber $k_c=5.50\:\mathrm{[rad/m]}$ with $15\%$ bandwidth. When propagating in a duct filled with ambient air, this carrier wavenumber corresponds to the carrier frequency $\omega_c=300\:\mathrm{[Hz]}$. The Gaussian pulse $p_g$ is truncated where the envelop falls 40dB below the peak.

Consider a temperature gradient $T_0(x)$ with $T_h=750\:\mathrm{[K]}$ being imposed on the REGs of the $N$-cell TAWG. As predicted by the eigenvalue analyses conducted earlier, this selected $T_h$ is below the onset temperature of the ITA1 mode in the five-cell TAWG, yet in between the onset temperature of the ITA1 and ITA2 modes in the ten-cell TAWG. Figures \ref{time}(b.1) and (b.2) show the space-time plots of the wave packet propagation in the five-cell ($N=5$) TAWG, when the wave packet impinges from both the left and the right hand sides of the TA section, respectively. The two red dashed lines represent the boundaries of the TA section. In each case, the initial wave packet, once released, splits into two wavefronts. The wavefront that (backward) propagates towards the end of the TAWG is absorbed by the anechoic termination, while the wavefront that (forward) propagates towards the TA section starts interacting with the $N$-cell TA section. Upon interaction, this wavefront gets partially reflected and partially transmitted, with the amplitude being amplified or attenuated depending on the TA coupling. The reflected and transmitted waves are eventually absorbed by the anechoic terminations on both sides. The TA coupling breaks the nonreciprocity of the waveguide, as already discussed in \cite{HaoPRB} and \cite{Olivier}. Therefore, the strength of the reflected or the transmitted waves is different in both cases, although the initial wave packet is symmetrically released on either side of the TA section. Clearly, if the wave packet propagates through the TA section in the direction of the rising temperature in the REG, as shown in Fig. \ref{time} (b.1) or (b.3), both the reflected and transmitted waves carry more energy than when the wave packet propagates in the opposite direction (Fig. \ref{time} (b.2) or (b.4)).

When the same wave packet is launched in the ten-cell ($N=10$) TAWG, similar phenomena take place, as shown in Figs. \ref{time}(b.3) and (b.4). However, regardless of the side of the TA section the initial wave packet impinges from, a wave with significantly different wavelength appears in the TAWG, after the initial transients are absorbed by the boundaries. The amplitude of this wave grows with time. Note that in Figs. \ref{time}(b.3) and (b.4), the color of the wave pattern becomes darker with time. Remember that $T_h=750\:\mathrm{[K]}$ exceeds the onset temperature of the ITA1 mode in the ten-cell TAWG, so this wave is expected to be the ITA1 mode being excited. Also note that the range of the color map is limited to $p\in[-0.5,0.5]\:\mathrm{[Pa]}$ in order to visualize the propagation of the initial wave packet. Any pressure fluctuation of the excited ITA1 mode that is higher than 0.5 [Pa] is conflated in the dark red region.

To further substantiate the appearance of the excited ITA1 mode, we probed the pressure fluctuation at the right end of the TA section (shown as the red dot in Fig. \ref{time}(a)); in this case, the wave packet impinges from the left, as shown in Fig. \ref{time}(a). The probed pressure is plotted in Figs. \ref{time}(c.1) and (c.2), respectively for the $N=5$ and $N=10$ configurations. Figure \ref{time}(c) is plotted in the log scale, so all pressure fluctuations with negative values are neglected. In the five-cell configuration, no ITA1 mode is excited after both the reflected and transmitted waves are absorbed. However, in the ten-cell configuration, a single-frequency oscillation emerges after approximately $t=0.15\:\mathrm{[s]}$, and grows with time exponentially. One can recover the angular frequency and the growth rate, or equivalently the negative decay rate, in Fig. \ref{time}(c.2), by calculating the peak-to-peak time duration (period $T=2\pi/\omega_r$) and the slope of the red dashed straight line (slope $s=-(\mathrm{log}_{10}e)\omega_i$), respectively. From Fig. \ref{time}(c.2), the eigenfrequency of the ITA1 mode is calculated as: $\omega/(2\pi)=\omega_r/(2\pi)+i\omega_i/(2\pi)=42.5532-7.2670i\:\mathrm{Hz}$, that is in excellent agreement with the result returned by the eigenvalue solver for the ITA1 mode, $\omega/(2\pi)=42.5920-7.2766i\:\mathrm{Hz}$. The numerical results summarized in Fig. \ref{time} provide clear evidence that with a sufficiently strong TA coupling, the ITA modes can be excited by the temperature gradient being imposed on the REGs. 

\section{Discussion}
Traditionally, TA instabilities appear in a finite-sized system. The TA coupling destabilizes the pre-existent acoustic modes, such as the standing-wave modes in a long duct terminated by either sound hard or sound soft boundary conditions. Therefore, the angular frequency $\omega_r$ of the unstable acoustic modes is close to the uncoupled angular frequency. In practice, the length $l_s$ of the TA coupling element, e.g. the REG, is much shorter than the total length $L$ of the whole device, so the angular frequency is primarily determined by the wavelength of the acoustic modes, or the total length $L$. However, for the ITA modes explored in this study, either the angular frequency $\omega_r$ or the decay (growth) rate $\omega_i$ is very sensitive to the temperature gradient, or equivalently the value of $T_h$ (see Figs. \ref{eigfreq}(a-c)). Remember that the existence of the ITA modes does not rely on any acoustic feedback provided by the boundary of the finite waveguide, so one may not be able to accurately estimate the frequency of the unstable ITA modes by considering solely the uncoupled structure. Especially noteworthy is the most unstable mode, which is the ITA1 mode. This mode degenerates to the overdamped AE1 mode when $T_h=300\:\mathrm{[K]}$, which has $\omega_r=0$. In practice, numerical simulations involving the complete TA coupling have to be conducted to determine the accurate frequency of the ITA modes, to facilitate an effective TA noise mitigation strategy. For example, a properly crafted Helmholtz resonator that targets the eigenfrequency of the unstable ITA mode can be leveraged for suppression of the TA noise \cite{Umut,Cora}.

Now, we revisit the two existing experimental studies of TAWGs, which were referred to as the TA diode \cite{Biwa2016} and the cascade TA amplifier \cite{Senga}, respectively. The TA diode \cite{Biwa2016} consisted of four unit cells, or equivalently, four REGs, which were connected to an inlet duct and an outlet duct. A loudspeaker was placed on the end of the inlet duct to generate harmonic waves, while the outlet duct was ended with an anechoic termination. The highest hot temperature $T_h$ reported in \cite{Biwa2016} is $T_h=690\:\mathrm{[K]}$. Yet, no TA instability was observed or recorded in the experiments. Based on our analyses in Section \ref{ITAmodes}, the absence of instability may be attributed to the fact that $T_h=690\:\mathrm{[K]}$ is below the onset temperature for the ITA modes of the four-cell system experimented in \cite{Biwa2016}. With the distinction between the selection of parameters in \cite{Biwa2016} and the present study acknowledged, we note that in the five-cell setup explored in this study, the onset temperature for the ITA1 mode is $T_{h_\mathrm{on}}=980\:\mathrm{[K]}$, higher than the selected $T_h$ used in the four-cell system in \cite{Biwa2016}. 

The TA amplifier investigated in \cite{Senga} consists of eight unit cells, in which the hot ends of the REGs were kept at $T_h=673\:\mathrm{[K]}$. However, the eight-cell device was terminated by a loudspeaker on each side. The loudspeakers were tuned so that the impedance at the ends of each unit cell was identical to ensure a harmonic oscillation at any point inside the amplifier. Such treatment artificially manipulated the reflection and transmission in each unit cell, which allowed a spatial amplification of the time harmonic wave. This phenomenon was well captured by the complex band structure analyses conducted in \cite{HaoPRB}. With the impedance boundary conditions being imposed by the loudspeakers (or Floquet periodicity in \cite{HaoPRB}), no unstable modes existed in the eight-unit amplifier. Nevertheless, terminating the finite periodic TAWG with well-tuned impedance may not always be feasible in practice, so specific TA instability suppression treatments have to be in place in the application of TAWGs.

Although the unstable ITA modes in the REG-based TAWG resembles their counterparts in the combustion systems (subject to flame-acoustic coupling), they are still quite different thermoacoustic instabilities, especially considering the large losses in the present TAWG system. The key contribution of this work lies in the comprehensive theoretical framework that is capable of capturing and explaining the evolution of the ITA modes from the AE modes upon increased TA coupling. The numerical modeling methodology adopted to predict the complex eigenfrequency of the ITA modes also benefits the practical design of effective instability mitigation treatments.

We have considered the ITA instability as a potential risk for the application of TAWGs, especially when they are designed for spatial amplification \cite{Senga}, non-reciprocal transmission \cite{Biwa2016, HaoPRB, Olivier}, or cloaking \cite{HaoPRB}. However, we believe that the understanding of the ITA mechanisms may also bring opportunities for further practical implementations, should the self-excited acoustic energy be properly leveraged. One significant difference between the ITA instability and its counterpart in a conventional TA engine lies in the fact that the modal frequency of the ITA instability is much more sensitive to the variation of temperature gradient than to the change of the actual device size. This aspect, as seen in Figs. \ref{eigfreq}(a-c), may lead to the design of tunable vibration sources (i.e., engines) that can output oscillatory motion of tunable frequencies just by controlling the temperature gradient applied to the TAWG. In this regard, the development of TA energy conversion devices may benefit from this new discovery both in terms of (1) additional power output flexibility, and (2) the potential for more compact devices. The ITA modes in TAWGs may also be implemented for the discovery of nonlinear acoustic waves, the most immediate being the generation of solitary waves in a straight waveguide, as a counterpart of the TA solitary wave in a looped tube \cite{Shimizu}. 

\section{Conclusions}
In this paper, we comprehensively analyzed the acoustic modes in a finite TAWG that is terminated by anechoic boundary conditions. In the absence of the spatial temperature gradient along the REGs, the abrupt cross-sectional area expansions at the REG ends allow the existence of the AE modes that have complex eigenfrequencies. The thermoviscous effects in the thin REG channels bring a unique set of eigenmodes, dubbed the TVE modes, that do not appear in the inviscid waveguide. The TVE modes have purely imaginary frequencies, yet purely real local wavenumber $k_d$ in the REG channels. Nevertheless, when the TA coupling (the spatial temperature gradients) is negligible, the acoustic modes (whether the AE or TVE modes) in the waveguide are evanescent modes that strongly decay with time, due to the large acoustic losses introduced by the anechoic terminations. 

However, with the increase of the temperature gradients, the AE modes evolve towards unstable modes, with the imaginary part of their eigenfrequencies turning from positive- to negative-valued. The appearance of these unstable TA modes, referred to as the ITA modes, are attributed to the intrinsic interaction between the thermal energy introduced by the spatial gradients and the pressure acoustic fields in the anechoically terminated system with no positive acoustic feedback. This work provides the first demonstration of the ITA modes in the REG-based TA systems, which possesses essential distinction from the unstable standing or traveling TA modes in traditional TA devices. The angular frequency of the ITA modes is highly sensitive to the hot temperature $T_h$ imposed on the REGs.

The existence of ITA modes hinders the practical applications of TAWGs, which posses several intriguing capabilities of sound manipulation, e.g., nonreciprocal and/or zero-index transmissions. However, the analytical and numerical framework introduced in this work is capable of accurately predicting the angular frequency and growth (decay) rate of each ITA mode, thus shedding lights on the design of effective instability suppression strategies. On the other hand, the extensive study ITA modes presented in this work may also shed light on the development of novel TAEs with enhanced tunability and compact dimensions. More importantly, the main analytical methodologies adopted in this study may have implications on addressing stability issues of active acoustic devices that leverage other multi-physics coupling elements, e.g. the electroacoustic devices.

\bibliography{ITA}

\end{document}